\documentclass[useAMS,usenatbib]{mn2e}

\usepackage{amssymb}
\usepackage{times}
\usepackage{graphicx}
\usepackage{subfigure}

\newcommand{\beq}{\begin{equation}}
\newcommand{\eeq}{\end{equation}}

\def\gs{\mathrel{\lower0.6ex\hbox{$\buildrel {\textstyle >}\over{\scriptstyle \sim}$}}}
\def\ls{\mathrel{\lower0.6ex\hbox{$\buildrel {\textstyle <}\over{\scriptstyle \sim}$}}}
\newcommand{\simgt}{\lower.5ex\hbox{$\; \buildrel > \over \sim \;$}}
\newcommand{\simlt}{\lower.5ex\hbox{$\; \buildrel < \over \sim \;$}}

\newcommand{\aap}{A\&A}
\newcommand{\apj}{ApJ}

\newcommand{\apjs}{ApJS}
\newcommand{\aj}{AJ}

\newcommand{\prd}{Phys. Rev. D}
\newcommand{\mnras}{MNRAS}
\newcommand{\physrep}{Phisycs Rep.}

\newcommand{\jcap}{JCAP}

\begin{document}

\title[Elastic dark energy]{Dark energy as an elastic strain fluid}
\author[N. Radicella et al.]{N. Radicella$^{1,2}$, M. Sereno$^{3,4}$\thanks{E-mail: mauro.sereno@polito.it (MS)}, A. Tartaglia$^{3,4}$
\\
$^1$Dipartimento di Fisica ``E.R.Caianiello", Universit\`a di Salerno, I-84084 Fisciano, Italia\\
$^2$INFN, Sezione di Napoli, GC di Salerno, I-84084 Fisciano, Italia\\
$^3$Dipartimento di Scienza Applicata e Tecnologia, Politecnico di Torino, corso Duca degli Abruzzi 24, I-10129 Torino, Italia\\
$^4$INFN, Sezione di Torino, via Pietro Giuria 1, I-10125, Torino, Italia
}


\maketitle

\begin{abstract}
The origin of the accelerated expansion of the universe is still unclear and new physics is needed on cosmological scales. We propose and test a novel interpretation of dark energy as originated by an elastic strain due to a cosmic defect in an otherwise Euclidean space-time. The strain modifies the expansion history of the universe. This new effective contribution tracks radiation at early times and mimics a cosmological constant at late times. The theory is tested against observations, from nucleosynthesis to the cosmic microwave background and formation and evolution of large scale structure to supernovae. Data are very well reproduced with Lam\'e parameters of the order of $10^{-52}\mathrm{m}^{-2}$.
\end{abstract}

\begin{keywords}
        cosmology: theory  -- dark energy
\end{keywords}

\section{Introduction}

The discovery of the accelerated expansion of the universe \citep{riess,perl} was rather a surprise and gave way to an intense theoretical quest for an explanation. At the cosmological scale, a new ingredient is indeed needed in order to drive the present acceleration. Referred to as dark energy, the nature and nurture of this new term is still debated with the cosmological constant  being the simplest candidate. Although added ad hoc, it provides the bases for the  $\Lambda$CDM model  that deserves the name of concordance model since it is able to fit extremely well the full available data set. Notwithstanding this remarkable success, the $\Lambda$CDM is theoretically unappealing because of several well known shortcomings. This fact motivates the search for other dark energy candidates with an appropriate evolution of the equation of state parameter $w$: somewhere its value must range from $-1/3$ to $-1$, or even less than $-1$ in the case of ``phantom energy", in order to produce an accelerated expansion.

Limiting our attention to the classical approach, we see that the different theories have different motivations, but generally speaking, though preserving mathematical consistency, hardly correspond to physical intuition. An attempt to build a theoretical paradigm based on familiar concepts of classical physics at the meso-scale is the Strained State Theory (SST) whose cosmological version is the Strained State Cosmology (SSC) or Cosmic Defect Theory (CDT), described in \cite{CQGTR} and refined and tested in \cite{CQGRST,signature}. Basically the idea is that the strain of a curved space-time, defined as it is in three-dimensional solids and in the elasticity theory, plays a role being a component of the energy content of the universe. The strain is with respect to a flat reference manifold with all geometrical symmetries, i.e. a Euclidean four-dimensional manifold. The use of a Euclidean reference is an evolution and improvement with respect to \cite{CQGTR} and \cite{rad+al11} where a Minkowski reference manifold was used instead. In fact the presence of the light cones indicates that a Minkowski manifold is not the maximally symmetric flat undeformed manifold \citep{signature}.

Once the strain has been introduced in the form of a symmetric tensor depending on the Lam\'{e} coefficients of space-time, the remark is that its presence implies a deformation energy density even in vacuo. The theory introduces this additional contribution in the Lagrangian density, moulding it on the analogous three-dimensional case. Matter/energy is then expressed, as usual, in the form of additional terms where matter fields minimally couple to the geometry via the metric tensor. In practice it is the additional ``elastic" term which plays the role of a dark fluid permeating the whole space-time and producing what, in our $(3+1)$-dimensional view, we read as the accelerated expansion of the universe. 

The Lagrangian we have introduced resembles the one of a massive gravity model \citep{hin12}, with a potential term, the strain energy density, added to the canonical kinetic one, the Ricci scalar of the metric $g_{\mu\nu}$. The formal analogy at the level of the action, however, does not lead to the same phenomenology, the main difference being the fact that the reference metric used in the SST in order to construct the strain energy density is Euclidian. 

Another peculiarity of the SSC comes again in analogy with three-dimensional elastic continua. We ascribe to the universe at large the Robertson-Walker (RW) symmetry, i.e. spacial isotropy and homogeneity. Though the RW symmetry appears to be quite natural its origin is not in the matter distribution, which rather is a consequence of it. In our ``elastic" continuum paradigm the global symmetry is fixed by a texture defect (the ``initial" singularity) just as it happens in ordinary solids when defects are present.

The purpose of the present paper is to test the theory on the constraints posed by the existing evidence at cosmic scale and/or high redshift. The paper is organized as follows. Section~\ref{sec_outl} resumes the essential formulae of the SST; Sect.~\ref{sec_effe} analyzes the consistency of the interpretation of the strain energy as a dark energy. Section~\ref{sec_eucl} discusses the effect of using a Euclidean reference manifold rather than a Minkowskian one. Section~\ref{sec_obse} evaluates in detail, in its eight subsections, the constraints coming from the available data. Section~\ref{sec_fore} comments on the expected outcomes from next generation observations. Finally, Sec.~\ref{sec_conc} contains a discussion of the whole work together with some conclusions.

\section{Outline of the Strained State Theory}
\label{sec_outl}

The main ingredient of the theory is the strain tensor $\epsilon_{\mu\nu}$ written in terms of the actual metric tensor $g_{\mu\nu}$ and the Euclidean metric tensor $E_{\mu\nu}$:
\beq\label{strain}
\epsilon_{\mu\nu}=\frac{1}{2}(g_{\mu\nu}-E_{\mu\nu}).
\eeq
The associated strain energy density is expressed in terms of two parameters, the Lam\'{e} coefficients of space-time, $\lambda$ and $\mu$:
\beq
W_s=\frac{1}{2}\lambda\epsilon^2+\mu\epsilon_{\alpha\beta}\epsilon^{\alpha\beta}.
\eeq
Here $\epsilon=\epsilon_{\alpha}^{\alpha}$ is the trace of the strain tensor.

According to the SST approach the Lagrangian density of  space-time in vacuo is written:
\begin{equation}\label{lagrangian}
\mathfrak{L}=(R+\frac{1}{2}\lambda\epsilon^2+\mu\epsilon_{\alpha\beta}\epsilon^{\alpha\beta})\sqrt{-g},
\end{equation}
where $R$ is the scalar curvature. From Eq.~(\ref{lagrangian}) and using Eq.~(\ref{strain}) we can obtain, by functionally varying with respect to the metric tensor $g_{\mu\nu}$, the energy momentum tensor of the strained space-time. It is:
\begin{eqnarray}
T_{(\mathrm{e})\mu\nu}	&	=	 &\frac{\lambda}{2}(\frac{1}{2}g_{\mu\nu}\epsilon^2+g_{\mu\nu}\epsilon-2\epsilon\epsilon_{\mu\nu})\sqrt{-g} \nonumber \\
			&	+	 &\mu(\epsilon_{\mu\nu}+\frac{1}{2}g_{\mu\nu}\epsilon^{\alpha\beta}\epsilon_{\alpha\beta}-2\epsilon_{\mu}^{\alpha}\epsilon_{\alpha\nu})\sqrt{-g}
\label{tensor}
\end{eqnarray}

It must be kept in mind that the tensor $E_{\mu\nu}$ appearing in Eq.~(\ref{strain}) is \emph{not} a metric tensor at all in the natural manifold (the only one which actually exists); it is a non-dynamical symmetric tensor, even though its interpretation within the paradigm of the SST is that it \emph{would} be the metric tensor if the manifold was totally unstrained.

The total degrees of freedom of the theory for the cosmological problem (before any specific symmetry is introduced) correspond to the ten independent elements of the metric tensor, as in standard general relativity. The definition of the strain tensor given in Eq.~(\ref{strain}) does not introduce any new physical degree of freedom. The freedom of choice of the coordinates, always referred to the natural manifold only (the actual space-time), leads of course to different explicit forms for the Euclidean tensor, but this fact can be seen as a gauge freedom not producing any consequence on the physical configuration of the natural manifold.

Looking at Eq.~(\ref{tensor}) we see that it corresponds, in the usual interpretation scheme, to a fluid whose energy density and pressure may directly be read out from $T_{(\mathrm{e})\mu\nu}$. We may indeed exploit both the idea of strain and of a fluid as far as we assume, at the cosmological scale, a Robertson-Walker symmetry i.e. global homogeneity and isotropy of space. Under this symmetry the strain tensor turns out to be diagonal with equal space-space components \citep{CQGRST}. Density and pressure reads, respectively,
\begin{eqnarray}
  \rho_{(\mathrm{e})} c^2 &=& T_{(\mathrm{e}) 0}^{0} = \frac{3}{4}\mu\frac{2\lambda+\mu}{\lambda+2\mu}\frac{(a^2+1)^2}{a^4} \label{presdens1}\\
  p_{(\mathrm{e})} &=& -T_{(\mathrm{e}) i}^{i} = -\frac{\mu}{4}\frac{2\lambda+\mu}{\lambda+2\mu}\frac{3a^4+2a^2-1}{a^4} \label{presdens2}
\end{eqnarray}
The index $i$ labels any of the space coordinates of a rectangular reference frame; no summation is assumed in this case.

In a fully consistent general relativistic description we should remark that our ``fluid" is nothing that allows for differential flow. It expresses a peculiar symmetry of space-time and, in four dimensions, it corresponds to a global equilibrium configuration of the Riemannian manifold endowed with the Robertson-Walker symmetry. 

If we add dust and radiation to the scenario and apply the energy condition we may work out the general form of the Hubble parameter, $H=\dot{a}/a$, for our Friedmann-Lemaitre-Robertson-Walker universe:
\begin{equation}
\label{hubble}
  H^2=B\left[1+\frac{(1+z)^2}{a_0^2}\right]^2+\frac{\kappa}{6}(1+z)^3[\rho_\mathrm{m0}+\rho_\mathrm{r0}(1+z)].
\end{equation}
We have used the shorthand notation 
$$
B=\frac{\mu}{4}\frac{2\lambda+\mu}{\lambda+2\mu}.
$$ 
$a_0$ is the present value of the scale factor of the universe $a$ and is a parameter of the theory; $\kappa$ is the usual coupling constant of Einstein's theory; $\rho_\mathrm{m0}$ and $\rho_\mathrm{r0}$ are the present average matter and radiation energy densities in the universe; $z$ is the redshift. 

The use of the Euclidean reference manifold is reflected in the $+$ sign appearing in the first square brackets of Eq.~(\ref{hubble}). The definition of the $B$ parameter in terms of the Lam\'{e} coefficients also differs from the definition in \cite{CQGTR} since there the lapse function was rigidly fixed to 1 as if the reference and the natural frames were totally uncorrelated. Here just one coordinate system is used for both manifolds, so that the gauge freedom holds only once and the lapse function is obtained from the Lagrangian. Eq.~(\ref{hubble}) will be mostly used for the comparison with the observations.

\section{Effective equation of state of the strain}
\label{sec_effe}

\begin{figure}
       \resizebox{\hsize}{!}{\includegraphics{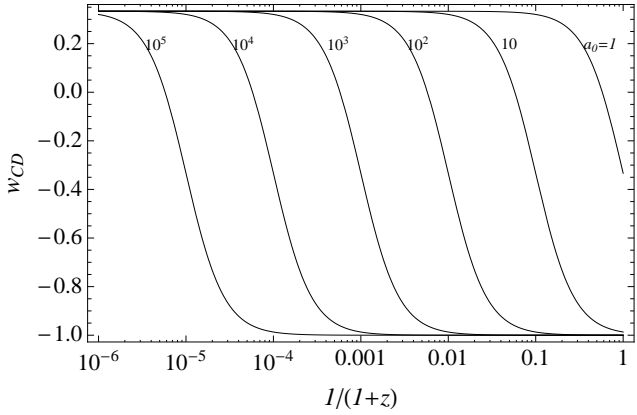}}
       \caption{Evolution with redshift of the effective equation of state of the elasticity strain for different values of $a_0$. The corresponding  value of $a_0$ is reported near each curve.}
	\label{fig_w_CD_vs_z}
\end{figure}

\begin{figure}
       \resizebox{\hsize}{!}{\includegraphics{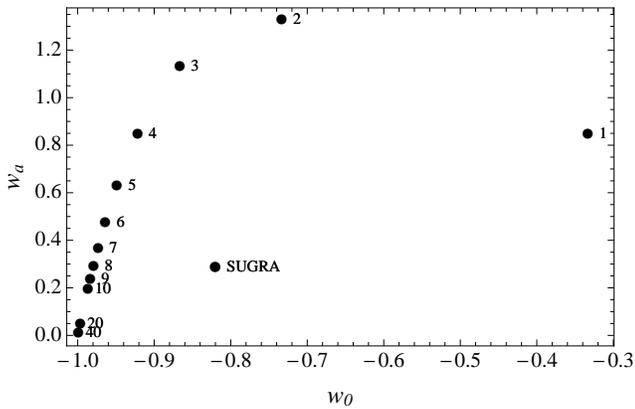}}
       \caption{Characteristic values of the effective equation of state of the elasticity strain for different values of $a_0$. Each point has the corresponding $a_0$ nearby. For comparison we also include values for alternative models. The point SUGRA denotes the supergravity model proposed in \citet{br+ma99}.}
	\label{fig_w_0_vs_w_a}
\end{figure}

\begin{figure}
       \resizebox{\hsize}{!}{\includegraphics{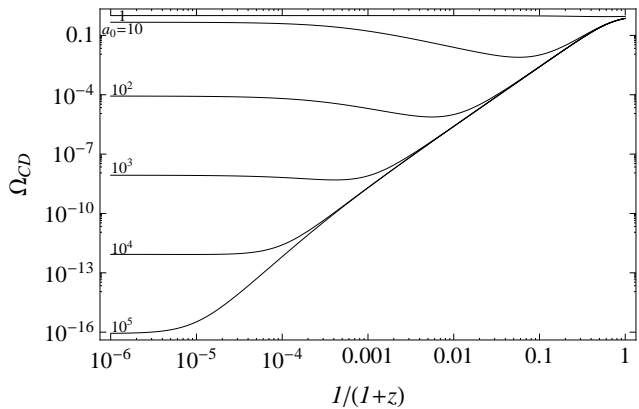}}
       \caption{Evolution with redshift of the effective energy density of the elasticity strain, in units of the critical density, for different values of $a_0$. The parameters $B$ and $\rho_\mathrm{m0}$ were fixed to the values in Tab.~\ref{tab_parameters}.}
	\label{fig_Omega_CD_vs_z}
\end{figure}

The SST provides a mechanism to set up an effective tracking dark energy. Sticking to the description of a strain fluid, we have it uniformly permeating the universe (in space) and evolving in cosmic time. We can deduce from Eqs.~(\ref{presdens1},~\ref{presdens2}) the equation of state. It is:
\begin{equation}
\label{state}
w_\mathrm{CD}=\frac{p_{(\mathrm{e})}}{\rho_{(\mathrm{e})}c^2}=-\frac{1}{3}\frac{3a^4+2a^2-1}{(a^2+1)^2}.
\end{equation}
There are two asymptotic behaviors, see Fig.~\ref{fig_w_CD_vs_z}. At early times ($a\ll1$), i. e., just outside the singular horizon from which the Lorentzian signature stems at $a=0$ \citep{signature}, the behavior of the strain ``fluid" is similar to radiation, $w_\mathrm{CD} \sim 1/3$. At late times ($a\gg1$), the SSC mimics a cosmological constant, $w_\mathrm{CD} \sim -1$. The transition from positive to negative pressure happens at $a=1/\sqrt{3}$. The corresponding redshift of the transition is then connected to the present value of the scale factor, $z_{w_\mathrm{CD}}=\sqrt{3}a_0-1$. The larger $a_0$, the earlier the transition, i.e., the larger the redshift. For $a_0\gg 1$, it is $z_{w_\mathrm{CD}}\simeq \sqrt{3}a_0$ and the transition $\Delta z_{w_\mathrm{CD}}$ takes nearly two logarithmic decades in redshift, i.e., starts at $\sim 10^1z_{w_\mathrm{CD}}$ and ends at $\sim 10^{-1}z_{w_\mathrm{CD}}$. In a single expression, $\log_{10}\Delta z_{w_\mathrm{CD}}/z_{w_\mathrm{CD}}\simeq 1$.

To compare the SSC with other dark energy/modified gravity models, we can parameterize the equation of state as $w(z) \simeq w_0+w_a z/(1+z)$, where $w_0$ is the present value of the equation of state and $w_a=2d w/d \ln (1+z)|_{z=1}$ \citep{lin03}. A cosmic defect acts like a dark energy whose equation of state is positively evolving, $w_a>0$, see Fig.~\ref{fig_w_0_vs_w_a}. The late time behavior of the effective equation of state deviates significantly from the cosmological constant for $a_0 \ls 10$, whereas is nearly indistinguishable from $\Lambda$ for $a_0 \gs 20$, when $w_0 \sim -1$ and $w_a \sim 0$.

Since the contribution to the expansion is initially radiation-like, the fraction of the total energy contributed by the defect mechanism can be significant at early times. The smaller $a_0$, the larger the early energy contribution, see Fig.~\ref{fig_Omega_CD_vs_z}. For $10\ls a_0\ls 100$, the cosmic defect contributes few percents of the total energy. By comparison, a cosmological constant has a fractional energy density at the $10^{-9}$ level at $z\sim 10^3$.

Even if tracking quintessence was introduced in the context of slow-rolling scalar fields \citep{ste+al99}, the SSC de facto shares the same desirable features. First, whatever the scale-length of the expansion factor, i.e., for every value of $a_0$, the effective energy from the defect may be significant today. Conditions in the early universe are related to $a_0$ but the late time contribution to the expansion is only sensitive to the Lam\'{e} coefficients. There is no ``coincidence problem". In the present version of the SSC, $a_0$ is not related to $\lambda$ and $\mu$. As far as $a_0 \gg 1$, a variation of the scale-length only affects the early time $\rho_\mathrm{(e)}$ whereas the today effective dark energy is only related to the Lam\'{e} coefficients through the factor $B$.

Secondly, when the universe is radiation-dominated ($a<<1$), then $w_\mathrm{CD} \sim 1/3$ and the effective energy density of the cosmic defect decreases as the radiation density. When the universe is matter-dominated  ($a \gg a_0 \rho_\mathrm{r0}/ \rho_\mathrm{m0}$), then $w_\mathrm{CD}$ is less than zero. At that time, $\rho_\mathrm{(e)}$ is nearly constant and decreases much less rapidly than the matter density.

\section{Euclidean vs Minkowskian background}
\label{sec_eucl}

\begin{figure}
       \resizebox{\hsize}{!}{\includegraphics{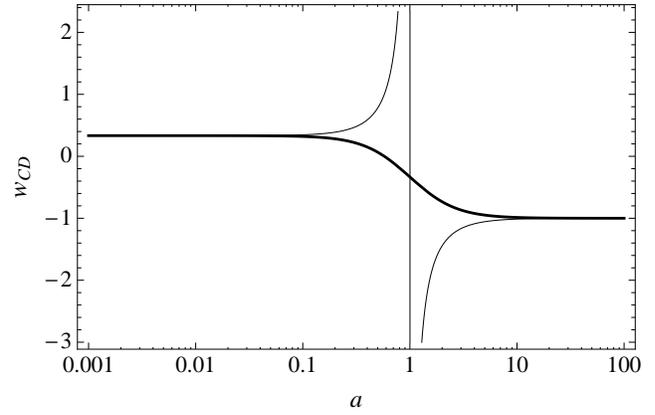}}
       \caption{Evolution of the effective equation of state of the elasticity strain for either an Euclidean (thick line) or Minkowskian (thin line) background as a function of the scale factor.}
	\label{fig_w_CD_Eucl_vs_Mink}
\end{figure}

\begin{figure}
       \resizebox{\hsize}{!}{\includegraphics{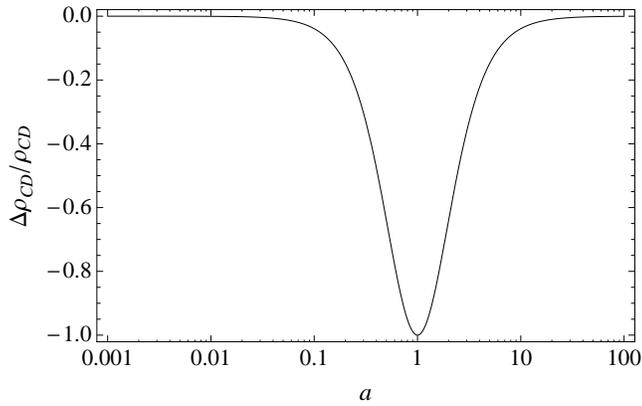}}
       \caption{Evolution of the relative difference of the effective density of the elasticity strain for a Minkowskian background with respect to an Euclidean background.}
	\label{fig_rho_CD_Eucl_vs_Mink}
\end{figure}

Besides theoretical a priori considerations which make an Euclidean reference system preferable to the Minkwskian alternative \citep{signature}, the deformation of the Euclidean background also brings a more regular evolution in. In fact, the Hubble parameter is a smooth function of $a$ for all real values of the scale factor excepting the initial singularity and there are no divergences in the evolution of the effective density and equation of state, see Figs.~\ref{fig_w_CD_Eucl_vs_Mink} and ~\ref{fig_rho_CD_Eucl_vs_Mink}. 

In the Minkowskian case a divergence appears at $a=1$. i.e., $z=a_0-1$. To distinguish on an observational ground the two cases, we would then need an accurate sampling of the expansion history of the universe at redshifts of the order of $z_{w_\mathrm{CD}}$. On the other hand, outside a small redshift range around $a \sim 1$, the two evolutions are very similar. 

However, some important differences are in order. The effective equation of state for the Euclidean case is bounded in the interval $-1<w_\mathrm{CD}<1/3$, whereas in the Minkowskian case $w_\mathrm{CD}$ approaches the asymptotic value of $-1$ from below after the divergence. The absence of divergences in the cosmological sector might be a hint in the direction of a good behavior of the theory when analyzing its propagating modes. In particular, the effect of an Euclidean reference manifold could cure the ghost problem and let the scalar sector behave as a healthy propagating mode.

Considering analogies and differences with massive theories of gravity \citep{hin12}, we know that the latter have six propagating modes: the five
propagating degrees of freedom for the massive spin 2 interaction and the so-called Boulware-Deser mode, avoided in the linear description of the theory
\citep{boulware} but reappearing at the non-linear level. The Boulware-Deser mode is always a ghost since, in the ADM formalism, the Hamiltonian of the
scalar mode, when written with a reference metric with Lorentzian signature, is not positive definite. Recently, \citet{hassan12a, hassan12b} have shown that in massive gravity theories carried out in the ADM formalism at the full non-linear level, there always exists a Hamiltonian constraint which eliminates the ghost with an associated secondary constraint. The sixth mode does not contribute to propagation either with a flat or a general reference metric but the condition for the metric to have Lorentzian signature explicitly comes into play in the derivation. In our case no problems arise for the cosmological solutions and our conjecture is that the Euclidean signature of the reference metric is what makes the difference, thus curing the problems also for the propagating perturbations.

\section{Observational constraints}
\label{sec_obse}

Having worked out the SST with an Euclidean reference both in the weak field, short range approximation \citep{CQGRST} and in the cosmological context, we are in position to compare the theory prediction with a comprehensive series of cosmological tests including kinematic expansion, formation and growth of the large scale structure, cosmic microwave background (CMB) and formation of nuclei. 

The main observational signature of the cosmic defect comes from its impact on the expansion of the universe. Local corrections are negligible \citep{CQGRST}, so that, in the SST, the gravitational potential felt by matter/radiation over-densities is de facto Newtonian. The growth of perturbations is then affected mainly through the modified expansion rate of the background. 

Different cosmological tests can probe the SSC in different redshift ranges. At low redshift, $z\ls 1$, the expansion rate as inferred from supernovae (SNe) measurements forces any dark energy model to approximate the cosmological constant, $w_0\sim -1$. However, we still have no direct observational constraints on the expansion rate of the universe at $z\gs 2$, so that probes exclusively sensitive to the universal expansion are severely limited in constraining the early behavior of dark energy. Any early feature must therefore be constrained by a combination of expansion rate and matter power spectrum measurements \citep{jo+ka11}. Constraints from primordial nucleosynthesis and from the CMB limit the early dark energy (EDE) fraction to be small, less than few percents, but not completely negligible \citep{wri07}.

Here, we summarize the observational tests we considered and discuss the results. Method and statistical approach are similar to \citet{rad+al11} with three main differences. Firstly, we study the SSC as developed on elastically deformed Euclidean background, differently from \citet{rad+al11} who considered a Minkowskian background. Secondly, we consider a more comprehensive sample of tests and observational data. Thirdly, we employ different priors for the parameters.

\subsection{Cosmic microwave background}


The temperature power spectrum of CMB probes the universe at the decoupling epoch, $z_\mathrm{LS} \sim 1090$, as well as the expansion history between now and the last scattering surface. One of the main signatures of the CMB is the acoustic scale of the spectrum, $l_\mathrm{A}$ \citep{hu+su96,kom+al10}. This scale can be expressed as
\beq
l_\mathrm{A}=(1+z_\mathrm{LS})\pi \frac{D_\mathrm{A}(z_\mathrm{LS})}{r_\mathrm{s}(z_\mathrm{LS})},
\eeq
where $r_\mathrm{s}$ is the sound horizon at recombination and $D_\mathrm{A}$ is the angular diameter distance to the last scattering surface. As any early dark energy shifts the sound horizon \citep{do+ro06,li+ro08}, the presence of a cosmic defect affects the location of the acoustic peaks, which depends on the expansion factor at the matter-radiation equality. \citet{kom+al10} determined the acoustic scale from the WMAP (Wilkinson Microwave Anisotropy Probe)-7 data. We consider $l_\mathrm{A} =302.69 \pm 0.76 \pm 1.00$, where the first error is the statistical error and the second error gives an estimate of the uncertainty connected to the model \citep{el+mu07}.

\subsection{Low redshift supernovae}

\citet{rie+al09} obtained an accurate measurement of $H_0$ from the magnitude--redshift relation of 240 low-$z$ Type Ia SNe at $z < 0.1$. They got $H_0=74.2 \pm 3.6~\mathrm{km}~\mathrm{s}^{-1}\mathrm{Mpc}^{-1}$.

\subsection{High redshift supernovae}

SNe of type Ia trace the late time expansion of the universe. We consider the sample for cosmological studies in \citet{kow+al08}, who measured the distance modulus $d(z)$ of 307 SNe,
\beq
\label{sne1}
d(z)=25+5\log_{10} \left[(1+z) \int_0^z \frac{(c/\mathrm{Mpc})}{H(z')}dz'\right].
\eeq

\subsection{Nucleosynthesis}

The strain energy affects the expansion rate at the nucleosynthesis whereas the cross-sections of nuclear interactions are not influenced \citep{rad+al11}. The result in \citet{ioc+al09} based on the abundance of light elements  can be rewritten as a stretch factor $X_\mathrm{Boost}=(1+\rho_\mathrm{DE}/\rho_\mathrm{r})= 1.025\pm 0.015$, where $\rho_\mathrm{DE}$ is an additional non standard contribution to the density/energy budget and $\rho_\mathrm{r}$ is the radiation-like energy density from standard species. 

\subsection{Large scale structure}

Perturbations can not grow in a universe expanding as a radiation-dominated background. Since its early contribution is in the form of a radiation-like term, the space-time strain postpones the matter dominance. The effective additional energy density provided by the strain affects the scale of the particle horizon at the equality epoch, which on turn is imprinted in the matter transfer function. From the final 2dF Galaxy Redshift Survey analysis \citep{col+al05}, we can take the result 
\beq
\label{lss1}
X_\mathrm{Boost}^{-1/2} \left( \Omega_\mathrm{m0}h\right) =  0.168 \pm 0.016,
\eeq
where $h$ is the Hubble constant $H_0$ in units of $100~\mathrm{km~s}^{-1}\mathrm{Mpc}^{-1}$ and $\Omega_\mathrm{m0}$ is the matter density in units of the critical density $\rho_\mathrm{cr}\equiv 3 H_0^2/(8\pi G)$.

\subsection{Linear growth}
The cosmic defect affects the expansion rate and the growth of perturbations is influenced due to friction. The equation for the growth $D$ is \citep{li+je03,ba+po12}
\beq
D^{''}+\frac{3}{2}\left( 1-\frac{w(a)}{1+X(a)}\right)\frac{D^{'}}{a}-\frac{3}{2}\frac{X(a)}{1+X(a)} \frac{D}{a^2}=0,
\eeq
where
\beq
X(a)=\Omega_\mathrm{m0}\left(\frac{a}{a_0}\right)^3\left( \frac{H_0}{H(\rho_{r0}=\rho_\mathrm{m0}=0)}\right)^2.
\eeq 
A prime denotes derivative with respect to $a$. The rate of structure growth, $f=d \ln D(a)/d \ln a$, was recently measured by \citet{toj+al12}, who considered a passively evolving population of galaxies. They measured the evolution of $f$ between $z = 0.25$ and $z = 0.65$ by combining data from the Sloan Digital Sky Survey (SDSS) I/II and SDSS-III surveys. We consider the measurements of $f$ as summarized in their table~1.

\subsection{Baryon acoustic oscillations}

\citet{per+al10} measured the baryon acoustic oscillations (BAO) exploiting the spectroscopic  SDSS Data Release 7 galaxy sample. They achieved a distance measure at redshift $z = 0.275$, of $r_s(z_\mathrm{d})/D_\mathrm{V}(0.275) = 0.1390 \pm0.0037$, where $r_\mathrm{s}(z_\mathrm{d})$ is the comoving sound horizon at the baryon-drag epoch, $D_\mathrm{V}(z) = [(1 + z)^2D^2_A c z/H(z)]^{1/3}$, where $D_A(z)$ is the angular diameter distance and $H(z)$ is the Hubble parameter. Since the power spectrum  was measured for different slices in redshift, they also found an almost independent constraint on the ratio of distances $D_\mathrm{V} (0.35)/D_\mathrm{V} (0.2) = 1.736 \pm 0.065$. We use both observational  constraints.

\subsection{Data analysis}

\begin{table}
\centering
\begin{tabular}{r@{$\,\pm\,$}lr@{$\,\pm\,$}lr@{$\,\pm\,$}l}
        \hline
        \noalign{\smallskip}
        	\multicolumn{6}{c}{Fitted parameters} \\
        	\multicolumn{2}{c}{$B$} &	\multicolumn{2}{c}{$\rho_\mathrm{m0}$} & \multicolumn{2}{c}{$B_{a_0}^{-1}$}      \\
         \multicolumn{2}{c}{[$10^{-52}\mathrm{m}^{-2}$]} & \multicolumn{2}{c}{$[10^{-29}\mathrm{g~cm^{-3}}]$} & \multicolumn{2}{c}{[$10^{52}\mathrm{m}^2$]}    \\
         2.24	(2.22)	&	0.05	&	0.252(0.255)	&0.007	&	0.004(0.009)	&0.005	\\
          \noalign{\smallskip}
          \hline
          	\multicolumn{6}{c}{Derived parameters} \\
	 	\multicolumn{2}{c}{} &	\multicolumn{2}{c}{$\Omega_\mathrm{m0}$} & \multicolumn{2}{c}{$a_0$} 	 \\
         	\multicolumn{2}{c}{} 	&	0.272	&	0.009	&	70		&	60	\\
      \hline
\end{tabular}
\caption{Results of the statistical analysis. $B$, $\rho_\mathrm{m0}$ and $B_{a_0}^{-1}$ are the parameters of the SSC used to fit the data. The distributions of the derived parameters have been obtained from those of the fitted parameters. Central locations and dispersions are computed as mean and variance of the posterior probability functions. The best fit values are reported in brackets.}
\label{tab_parameters}
\end{table}

We performed the statistical analysis with standard Bayesian methods \citep{le+br02,mac03}. The method is similar to \citet{rad+al11}, with some differences. We considered the parameter space spanned by the matter density $\rho_\mathrm{m0}$, the relativistic energy density, which is frozen at $\rho_\mathrm{r0} \simeq 7.8\times10^{-34}~\mathrm{g/cm}^3$, the factor $B$ which combines the Lam\'{e} parameters and a last parameter for the size of the scale factor, whose present value $a_0$ is described in terms of $B_{a_0}^{-1}$, with
\beq
\label{nucl2}
B_{a_0}\equiv \frac{8}{9}\kappa \rho_\mathrm{r0}a_0^4.
\eeq

Employed priors differ from \citet{rad+al11}. As a priori information for the scale factor $a_0$, we considered a distribution uniform in logarithmically spaced decades, as appropriate for parameters with only lower bounds. For the other parameters, we considered uniform priors. 

The parameter space was explored with standard Monte Carlo Markov chains methods. Results are summarized in Table~\ref{tab_parameters}. Together with the parameters used to describe the model, we list also results for some other quantities of cosmological interest whose distribution was derived from those of the fitted parameters.

The SSC provides an excellent fit to the data. We retrieved a total $\chi^2 \simeq 320.1$ for the best fit parameters for 314 degrees of freedom. The accuracy of the fit is slightly better than that of the flat $\Lambda$CDM model. Assuming a model of universe with a cosmological constant and zero curvature, we found $\chi^2 \simeq 322.7$ for $\Omega_\mathrm{M0}\simeq0.27$ and $H_0 \simeq 70.3~\mathrm{km~s}^{-1}\mathrm{Mpc}$ (315 degrees of freedom). 

Given the low number of free parameters and the comparable $\chi^2$ values, we can not prefer one model to the other one on a statistical basis. The small difference in the $\chi^2$ values is mainly due to the nucleosynthesis constraint. At late times the SSC drives an expansion indistinguishable from the $\Lambda$CDM model. Both models give excellent fits to the SNe and other low-redshift observations. On the other hand, the abundance of light elements is compatible with some additional radiation-like contribution to the total energy budget. In the SSC, this is provided by the defect which can boost the expansion early when the contribution from the elastic strain to the expansion, i.e, the early effective density $\Omega_\mathrm{CD}$, is proportional to $B/a_0^4$. The value of the present day scale factor $a_0$ must then be large enough to make the additional push to the expansion compatible with the bound from the nucleosynthesis. 

An early boost could be provided in the $\Lambda$CDM model by some extra relativistic species. As we will discuss in the next section, experiments at large redshifts could then separate the various competing models.

The $B$ parameter is tightly constrained, which in turn fixes the Lam\'{e} coefficients. $\mu$ must be of the order of $10^{-52}\mathrm{m}^{-2}$, either if $\mu \sim \lambda$ or not. On a theoretical ground, we expect a priori the two Lam\'{e} coefficients to be of the same order, so the constraint on $\mu$ can be read as a constraint on $\lambda$ too. The Lam\'{e} coefficients are then of the same order of magnitude as the cosmological constant in the popular $\Lambda$CDM model. This can be viewed as a further support for the elastic origin of the dark energy accelerating the universe expansion.

The matter density parameter is very well determined too. The required amount of dark matter is in line with what needed in the $\Lambda$CDM model and with observations on the scale of both galaxies and galaxy clusters.

The exploited data-sets do not provide tight constraints on the size of the scale factor. However, we know from observations that the boost to the expansion at early times due to the cosmic defect has to be small. $a_0$ has then to be small enough to let nucleosynthesis and large scale formation happen. Even if the scale factor is poorly determined, we can set a lower bound. We found that $a_0\gs 20$ at the 99.73 per cent confidence level. 

The fitted SSC model predicts a transition redshift from decelerated to accelerated expansion at $z_\mathrm{T}=0.75\pm0.03$, in agreement with observational constraints from SNe \citep[ see table 1]{cun09}. After the transition, the effect of the cosmic defect on the expansion of the universe is indistinguishable from a cosmological constant.

\section{Forecast}
\label{sec_fore}

\begin{figure}
       \resizebox{\hsize}{!}{\includegraphics{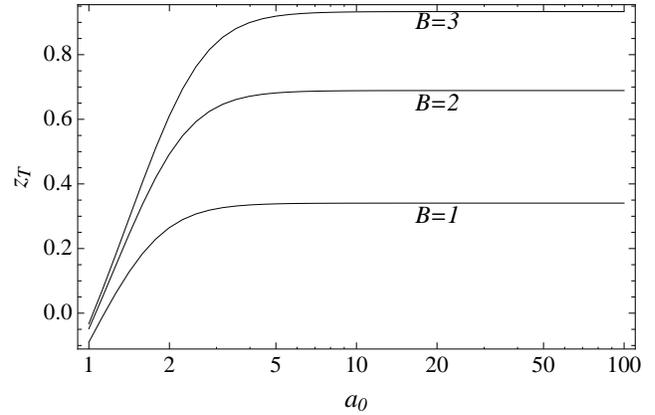}}
       \caption{Transition redshift as a function of the scale factor $a_0$ for different $B$'s. $B$ values are in units of $10^{-52}\mathrm{m}^{-2}$. The today matter (radiation) density is fixed to $0.25\times10^{-29}\mathrm{g~cm^{-3}}$ ($0.78\times10^{-33}\mathrm{g~cm^{-3}}$).
       }
	\label{fig_z_T_vs_a_0}
\end{figure}

Future and ongoing experiments promise to further improve observational constraints. The Planck satellite has been providing the first data  on the CMB, whereas the next generation of galaxy survey will exploit facilities such as the ground-based Large Synoptic Survey Telescope (LSST). The combination of the two probes will tighten what we know about the universal expansion rate and the matter perturbation growth \citep{jo+ka11}. 

As we have seen before, in the relevant redshift range covered by experiments, the SSC acts as an effective dark energy with a non-negligible fraction at high redshifts. We can then translate results on EDE \citep{jo+ka11} to forecast future bounds on the parameters of the SSC.

Results from Planck combined with a ground-based LSST-like survey should significantly improve present accuracy. A combination of weak lensing tomography, galaxy tomography, SNe, and the CMB should constrain the EDE density to 0.2 per cent of the critical density of the universe \citep{jo+ka11}.  

Since the parameter $B$ is very well constrained by late expansion as measured with observations of SNe, the constraint on the early effective density at large redshifts, when $\Omega_\mathrm{CD}$ is proportional to $B/a_0^4$, can then determine the scale factor $a_0$ to an accuracy of $0.05$ per cent.

A small fraction of EDE at the $\ls1$ per cent level can affect the formation of massive structures and may favor the early onset of star and galaxy formation. It can also explain the high level of Sunyaev-ZelÕdovich effect contribution to the high multipoles of the CMB temperature power spectrum \citep{li+ro08}. These tests could further probe the SSC.

An alternative way to improve the accuracy on the scale factor would be a very accurate determination of the transition redshift. However, $z_\mathrm{T}$ is very sensitive to the scale factor only for $a_0\ls 3$, see Fig.~\ref{fig_z_T_vs_a_0}, which is confidently excluded by already available data.

State of the art cosmological tests do not allow to distinguish between the Euclidean and the Minkowskian background. The evolution in the two cases differs in a redshift interval around $z=a_0-1$. Since we already know that $a_0 \gs 20$, the redshift range sampled by SNe ($z \ls 2$) can not disfavor any scenario. The same argument holds for other proposed standard sirens, such as coalescing massive black hole binaries emitting gravitational waves, which might be detectable out to $z \ls 10$-$15$  by next generation space-based observatories \citep{ses+al07,ser+al10}, or standard candles, such as gamma ray bursts detectable out to $z \ls 5$-$10$ \citep{ghi+al06}

The best chance to observationally favor one of the two hypotheses is for $a_0 \sim 10^3$, when differences show up at the formation of the CMB and would be detectable by future experiments. 

The additional degrees of freedom of the SSC or other EDE models degrade our ability to study neutrinos and other relativistic species with cosmological tests. The accuracy on the sum of neutrino masses by the future experiments discussed before is worsened by a factor two compared to the case without allowing for early DE \citep{jo+ka11}.

\section{Conclusions}
\label{sec_conc}

The SSC shares some interesting features with models of early dark energy and tracking quintessence which deserve interest on both the theoretical and the observational front. Due to friction in the expansion rate connected to the presence of dark energy, more structure had to form at earlier times than for a universe without dark energy in order to produce the matter perturbations seen in the present universe. The effect is even stronger in a universe with a non-vanishing amount of dark energy at early times, when more structure is required to have formed at earlier times than for a universe with only late-time dark energy. 

Most popular models behind EDE rely on tracking quintessence. The SSC shares the main advantages of the tracker solutions despite a very different context. While the latter alleviate the coincidence problem by considering a dynamical scalar field with a potential that brings the field evolution onto an attractor trajectory, in the SSC the modified expansion originates from a defect and the related elastic strain. The early contribution to the expansion is radiation-like and may be a significant fraction of the matter density during the matter dominated era, including the recombination epoch. At late times the strain acts as a cosmological constant. This behavior is compatible with a very large set of initial conditions.

Theories of gravity that deviate from general relativity at large distances may require some kind of screening on smaller scales \citep{cli+al12}. The deviation has to be sizable on cosmological scales but has to be suppressed down to at least five orders with respect to the usual Newtonian contribution on Solar System scales. The often invoked Vainshtein mechanism \citep{cli+al12} prescribes that higher order interactions suppress the extra modes near the local source of gravity and fields interact so strongly that they are frozen together and are unable to propagate freely. Other methods, such as either the chameleon \citep{ell+al89} or the symmetron \citep{hi+kh10}, may exploit the dependence of the effective potential on the environment.

The SST is intrinsically free from this problem. Cosmological observations strictly constrain the values of the Lam\'{e} coefficients ($\lambda \sim \mu \sim 10^{-52}\mathrm{m}^{-2}$). When considering the local gravitational field in the SST \citep{CQGRST}, an elastic strain of this size brings about a deviation suppressed by 20 orders of magnitude on the scale of the Solar System. The screening effect is then a straight consequence of the small values of the Lam\'{e} coefficients.

The setback of this might be some sort of fine-tuning problem similar to that affecting the cosmological constant. However, we can point out some substantial differences, that strongly mitigate the fine-tuning without referring to any debatable anthropic principle. First, the strain is not connected to any vacuum energy property, so we have no a priori guess on the values of the Lam\'{e} coefficients. Second, the discussion of the effective equation of state showed some sort of tracking mechanism. Whatever the value of $\lambda$ or $\mu$, the effective energy density is set to track a radiation-like energy at early times and to mimic a cosmological constant at late times.



\setlength{\bibhang}{2.0em}

\end{document}